\documentclass[10pt]{bmc_article}    

\usepackage{cite} 
\usepackage{url}  
\usepackage{ifthen}  
\usepackage{multicol}   
\usepackage[utf8]{inputenc} 
\usepackage{graphicx}
\usepackage{amsmath}
\usepackage{amssymb}
\usepackage{amsfonts}

\newboolean{publ}
\def\R{\mathcal{R}}

\newcommand{\PP}{{\mathbb P}}

\newenvironment{bmcformat}{\begin{raggedright}\baselineskip20pt\sloppy\setboolean{publ}{false}}{\end{raggedright}\baselineskip20pt\sloppy}

\begin{document}
\begin{bmcformat}

\title {A formal model of autocatalytic sets emerging in an RNA replicator system}
\author{Wim Hordijk\correspondingauthor$^1$%
         \email{Wim Hordijk\correspondingauthor - wim@WorldWideWanderings.net}
       and 
         Mike Steel$^2$%
         \email{Mike Steel - mike.steel@canterbury.ac.nz}%
       }
\address{%
    \iid(1)SmartAnalytiX.com\\
    \iid(2)Biomathematics Research Centre, University of Canterbury, Christchurch, New Zealand
}%
\maketitle

\begin{abstract}
{\bf Background:} The idea that autocatalytic sets played an important role in the origin of life is not new. However, the likelihood of autocatalytic sets emerging spontaneously has long been debated. Recently, progress has been made along two different lines. Experimental results have shown that autocatalytic sets can indeed emerge in real chemical systems, and theoretical work has shown that the existence of such self-sustaining sets is highly likely in formal models of chemical systems. Here, we take a first step towards merging these two lines of work by constructing and investigating a formal model of a real chemical system of RNA replicators exhibiting autocatalytic sets.\\
{\bf Results:} We show that the formal model accurately reproduces recent experimental results on an RNA replicator system, in particular how the system goes through a sequence of larger and larger autocatalytic sets, and how a cooperative (autocatalytic) system can outcompete an equivalent selfish system. Moreover, the model provides additional insights that could not be obtained from experiments alone, and it suggests several experimentally testable hypotheses.\\
{\bf Conclusions:} Given these additional insights and predictions, the modeling framework provides a better and more detailed understanding of the nature of chemical systems in general and the emergence of autocatalytic sets in particular. This provides an important first step in combining experimental and theoretical work on autocatalytic sets in the context of the orgin of life.
\end{abstract}

\ifthenelse{\boolean{publ}}{\begin{multicols}{2}}{}

\section*{Background}

Recently, significant new experimental results on spontaneous network formation among cooperative RNA replicators were reported \cite{Vaidya:12}. These results continue and strengthen a line of ongoing work on creating autocatalytic sets in real chemical systems \cite{Sievers:94,Ashkenasy:04,Hayden:08,Taran:10}. Moreover, they show the plausibility and viability of the idea of autocatalytic sets, especially in the context of the origin of life, as already developed in various forms several decades ago \cite{Kauffman:71,Kauffman:86,Kauffman:93,Eigen:77,Dyson:82,Rosen:91,Ganti:97}.

However, such chemical experiments, important as they are, are difficult, costly, and time-consuming to perform. In contrast, in our own work we have developed a theoretical framework of autocatalytic sets, which we consider a necessary condition for the origin of life, that can be studied computationally and mathematically \cite{Steel:00,Hordijk:04,Mossel:05,Hordijk:10,Hordijk:11,Hordijk:12,Hordijk:12a,Hordijk:12b}. This framework has provided many important insights into the emergence and structure of autocatalytic sets in its own right, and is considered to provide theoretical support for experimental observations \cite{Martin:07,Vaidya:12}.

The formal framework has, so far, mostly been applied to an abstract model of a chemical reaction system known as the binary polymer model. Even though this model already has a fair amount of chemical realism (for example, some recent experiments are an almost literal chemical implementation of the binary polymer model\cite{Taran:10}), a direct application to a real chemical system was still lacking. Here, we construct and analyze a formal model version of the recent experimental RNA replicator system \cite{Vaidya:12}, and show how this results in:
\begin{itemize}
  \item an accurate reproduction of experimental results,
  \item the correction of a misinterpretation of some of the original results,
  \item additional results and insights that could not be obtained from the experiments alone, and
  \item testable predictions about the behavior of the chemical system.
\end{itemize}
With this, we claim that the formal framework can be applied directly and meaningfully to real chemical systems, generating additional insights and predictions that are hard to obtain from experiments alone, thus providing a better understanding of real (bio)chemical systems. This represents an important first step towards merging experimental and theoretical work on autocatalytic sets.

\subsection*{Chemical reaction systems and autocatalytic sets}

We begin by briefly reviewing the relevant definitions and main results of the formal framework. A {\it chemical reaction system} (CRS) is defined as a tuple $Q=\{X,\R,C\}$ consisting of a set of molecule types $X$, a set of chemical reactions $\R$, and a catalysis set $C$ indicating which molecule types catalyze which reactions. We also consider the notion of a food set $F \subset X$, which is a subset of molecule types that are assumed to be freely available from the environment (i.e., they do not necessarily have to be produced by any of the reactions). Informally, an {\it autocatalytic set} (or RAF set) is now defined as a subset $\R' \subseteq \R$ of reactions (and associated molecule types) which is:
\begin{enumerate}
\item {\it reflexively autocatalytic} (RA): each reaction $r \in \R'$ is catalyzed by at least one molecule type involved in $\R'$, and
\item {\it food-generated} (F): all reactants in $\R'$ can be created from the food set $F$ by using a series of reactions only from $\R'$ itself.
\end{enumerate}

A more formal (mathematical) definition of RAF sets is provided in \cite{Hordijk:04,Hordijk:11}, including an efficient algorithm for finding RAF sets in a general CRS. It was shown (using the binary polymer model) that RAF sets are highly likely to exist, even for very moderate levels of catalysis (between one and two reactions catalyzed per molecule, on average) \cite{Hordijk:04,Mossel:05,Hordijk:10}, and that this result still holds when a more realistic ``template-based'' form of catalysis is used \cite{Hordijk:11,Hordijk:12}. 

The RAF sets that are found by the RAF algorithm are called {\it maximal} RAF sets (maxRAFs). However, it was shown that a maxRAF can often be decomposed into several smaller subsets which themselves are RAF sets (subRAFs) \cite{Hordijk:12a}. If such a subRAF cannot be reduced any further without losing the RAF property, it is referred to it as an {\it irreducible} RAF (irrRAF). The existence of multiple autocatalytic subsets can actually give rise to an evolutionary process \cite{Vasas:12}, and the emergence of larger and larger autocatalytic sets over time \cite{Hordijk:12a}.

\subsection*{A model of the RNA replicator system}

The main idea behind the RNA replicator system reported recently \cite{Vaidya:12} is the assembly (ligation) of ribozymes (catalytic RNA molecules) from two smaller RNA fragments. These ribozymes can then catalyze the assembly of other ribozymes (or in some cases their own assembly). Which ribozymes catalyze which assembly reactions is determined by one specific nucleotide in the ``guide sequence'' of the (potential) catalyst and one other specific nucleotide in the ``target sequence'' of a reactant. If these two nucleotides are each other's base pair complement, then the given ribozyme can catalyze the given reaction. Starting with RNA fragments with different nucleotides in these guide and target sequences, a mixture of auto- and cross-catalytic RNA replicators evolves over time (by means of the assembly reactions taking place), in which cooperative networks (autocatalytic sets) form spontaneously \cite{Vaidya:12}.

The ribozymes in this system are labeled M$j$N, where M denotes the specific nucleotide (A, C, G, or U) in the guide sequence of an RNA molecule, and N denotes the specific nucleotide in its target sequence \cite{Vaidya:12}. The value of $j$ denotes the specific location where the ribozyme was assembled from two smaller RNA fragments (there are three possible locations in the original experiments). However, as in some of the experimental results \cite{Vaidya:12}, for the purposes of looking for autocatalytic sets, this value can be ignored. So, in total there are $4\times4=16$ possible ribozymes M$j$N. For example, G$j$A and A$j$C are two such ribozymes.

The reaction set $X$ in the formal model of this RNA replicator system contains these 16 M$j$N ribozymes plus the smaller RNA fragments from which they are assembled. In fact, these RNA fragments are the subset of molecules that forms the food set $F$. In the model, we simply lump these fragments together into just two food molecule types (i.e., each ribozyme is the assembly of two ``generic'' fragments). For most of the results shown here this suffices, although a refinement will be made later on in one of the computational experiments (below).

The reaction set $\R$ in the model consists of the 16 assembly reactions that create the ribozymes from food molecules. For simplicity, the notation M$j$N will denote both a molecule type (ribozyme) as well as the reaction that created it. Finally, the catalysis set $C$ consists of all molecule/reaction pairs where the nucleotide M in the guide sequence of a molecule is the base-pair complement of the nucleotide N in the target sequence of a reaction. For example, molecule G$j$A can catalyze the reaction that produces molecule A$j$C, since G and C are complementary. The catalysis set $C$ thus consists of all such matching pairs ($16 \times 4 = 64$ in total).

The full mathematical definition of the CRS $Q=\{X,\R,C\}$ of the RNA replicator system is thus as follows:
\begin{itemize}
  \item $F$ = \{$f_1$,$f_2$\}
  \item $X$ = $F$ $\cup$ \{M$j$N $|$ M,N $\in$ \{A,C,G,U\}\}
  \item $\R$ = \{M$j$N $|$ M,N $\in$ \{A,C,G,U\}\}
  \item $C$ = \{(M$_1$$j$N$_1$, M$_2$$j$N$_2$) $|$ M$_i$,N$_i$ $\in$
        \{A,C,G,U\}, $i \in \{1,2\}$, and (M$_1$,N$_2$) $\in$
        \{(A,U),(C,G),(G,C),(U,A)\}\}
\end{itemize}
Note that since the reactants in all 16 reactions in $\R$ are food molecules, every subset $\R' \subseteq \R$ is automatically food-generated (F). Therefore, identifying possible RAF sets in this system only requires checking the (RA) part of the definition. We now analyze this model in detail and compare the results with those from the original experiments \cite{Vaidya:12}.

\section*{Results and discussion}

\subsection*{The existence of RAF sets}

Applying the RAF algorithm to the reaction set $\R$ (as defined above), returns the set $\R$ itself. In other words, the system  as a whole (all 16 reactions) forms a maximal RAF set. Moreover, this would be expected, given the extent of catalysis, even if the actual assignment of catalysis was randomized. In fact, under such a (random) model the probability that $\R$ forms an RAF is approximately $(1-e^{-4})^{16} = 74\%$. This can be derived as follows.

For molecule $x$ and reaction $r$, let  $I(x,r)$ be the indicator function defined by:
$$I(x,r)=
  \begin{cases}
  1, & \mbox{ if } x \mbox{ catalyzes }  r; \\
  0, & \mbox{ otherwise. }
  \end{cases}
$$
Now suppose that $Q = (X,\R, C)$ is a catalytic reaction system. Then under the random model \cite{Steel:00,Hordijk:04,Mossel:05}, the collection of indicator functions $(I(x,r): x \in X, r \in \R)$  are independent and identically distributed Bernoulli random variables. Thus, if we let $p = \PP [I(x,r)=1]$, then the probability $q_r$ that a given reaction $r\in \R$ is catalyzed by at least one molecule from $X$ is:
\begin{equation}
\label{eq1}
  q_r = 1-(1-p)^{|X|}.
\end{equation}
Moreover, if we set the expected amount of catalysis in the random model to the value observed in $Q$ then:
\begin{equation}
\label{eq2}
  p= \frac{|C|}{|X|\cdot |\R|}.
\end{equation}
Thus, Eqn. (\ref{eq1}) becomes:
\begin{equation}
\label{eq3} 
  q_r = 1-(1-p)^{|X|} \simeq  1- \exp(-|C|/|\R|).
\end{equation}
Now, if $\R$ is F-generated, and if $X$ is the set of molecules involved (as products or reactants) in $\R$, then $\R$ is an RAF precisely if every reaction in $\R$ is catalyzed by at least one molecule from $X$. This latter probability, under the random model, is:
\begin{equation}
\label{eq4} 
  \prod_{r \in \R} q_r =\left (1- \exp\left(-\frac{|C|}{|\R|}\right)\right)^{|\R|}
\end{equation}
which, for $|C|= 64$ and $|\R|= 16$ (as in the RNA replicator model) evaluates to $(1-e^{-4})^{16}$, as claimed. Notice that Eqn. (\ref{eq4}) is essentially independent of $|X|$ (the number of molecules) provided this is not too small.

Even though the entire reaction set $\R$ is a maximal RAF set in itself, RAF sets can often be decomposed into smaller RAF subsets \cite{Hordijk:12a}. Two such subsets were presented in the original experimental results (\cite{Vaidya:12} Fig. 4). However, a comparison of those diagrams with the formal model indicates that they are incomplete, as several catalysis arrows are missing. Fig. 1 (here) shows the corrected diagrams, with all catalysis arrows included. The first corrected diagram (Fig. 1, left) contains an RAF set of size seven (enclosed within the box): each node (reaction) is catalyzed by at least one of the members of the set, which satisfies the (RA) part (and, as noted above, the (F) part is automatically satisfied). Thus, the conclusion from the original experiments that ``{\it at the 1h time point, no closed network was possible}''\cite{Vaidya:12} is a misinterpretation due to the omission of several catalysis arrows. As the corrected diagram shows, at the 1h time point an RAF set of size seven already exists (Fig. 1, left), which has grown to size 11 at the 8h time point (Fig. 1, right, which forms an 11-reaction RAF set).

\subsection*{The structure of RAF sets}

As mentioned, RAF sets can often be decomposed into smaller subRAFs. Indeed, even the 7-reaction RAF subset in Fig. 1 (left) itself is composed of many smaller subRAFs. Previously we introduced a formal method to identify and classify RAF subsets \cite{Hordijk:12a}, resulting in a so-called {\it Hasse diagram} that visualizes the partially ordered set of all possible subRAFs and their subset relations. Applying this method to the 7-reaction RAF set in Fig. 1 (left) results in the Hasse diagram shown in Fig. 2.

This Hasse diagram contains 68 nodes, i.e., there are 68 possible subRAFs in the given 7-reaction RAF set. Note that there are $2^7 = 128$ possible subsets of a set of seven elements. So, more than half of these actually form RAF sets themselves, which shows how ``rich'' and diverse the RNA replicator system really is. In fact, if the same ratio (about half) holds for the full 16-reaction maximal RAF set, then we can expect more than $2^{16}/2 \approx 32,000$ nodes (subRAFs) in the Hasse diagram of the maxRAF, i.e., far too many to visualize in a meaningful way.

The edges in the Hasse diagram of an RAF set show the many possible ways in which the full RAF set can be built up from smaller subsets, or, in other words, how RAF sets can emerge and evolve \cite{Vasas:12,Hordijk:12a}. Which of these trajectories is actually followed depends on, for example, initial conditions and stochastic events such as ``spontaneous'' (uncatalyzed) reactions. As shown above, in the original experiment the system went through a stage of a particular 7-reaction RAF set, which then grew to an 11-reaction RAF set. However, what the Hasse diagram suggests, given the many possible subRAFs and ways of combining them into larger RAFs, is that when the experiment is repeated, most likely a {\it different} trajectory will be followed. This is a testable prediction that follows directly from the formal model and its results.

\subsection*{The emergence of RAF sets}

Performing the actual chemical experiments in a laboratory is costly and time-consuming. However, we can use the formal model to {\it simulate} molecular flow on the reaction network \cite{Hordijk:12b}. Using the well-known Gillespie algorithm \cite{Gillespie:76,Gillespie:77}, we performed such simulations on the full 16-reaction model, starting with an initial supply of food molecules (RNA fragments) only. Fig. 3 shows the result of one such simulation (time and concentration are in arbitrary units). For simplicity, we set all reaction rates equal, but with a factor $c$ difference between catalyzed and uncatalyzed (``spontaneous'') reactions. We tried various values for this factor $c$ (anywhere from $c=2$ to $c=100$), but qualitatively there is little difference in the overall dynamics, except that everything happens at longer or shorter time-scales, depending on the exact value of $c$.

As the figure shows, over time, all 16 M$j$N molecules (ribozymes) are produced, but exist in different concentrations at any given time point. This is also reflected in the different sizes of the nodes in the original experimental results (\cite{Vaidya:12} Fig. 4). However, over different runs of the simulation, this distribution of concentrations varies. Moreover, the order in which the 16 M$j$N molecules come into existence differs between simulations as well, as do the various (larger and larger) subRAFs the system goes through. For example, in the simulation shown in Fig. 3, the system goes through the sequence of subRAFs shown in Fig. 4. It starts with a subRAF of size 1 (Fig. 4, top left), which grows to size 3 (top right) and 7 (bottom left), then goes through sizes 8, 9, and 10 (not shown), then size 11 (bottom right), and eventually reaches the full 16-reaction maximal RAF set. Note, however, that the 7-reaction and 11-reaction subRAFs are different from those observed in the original experimental results (Fig. 1, and \cite{Vaidya:12} Fig. 4). Indeed, the system has gone through a different trajectory in the simulation run compared to the chemical experiment.

With these results we do not intend to claim that the simulation is an exact reproduction of the original experiment, where at regular intervals 10\% of the current solution was transferred to a new solution of fresh RNA fragments \cite{Vaidya:12}. Although we can include such ``transfer'' steps in our simulations as well, we mainly wish to show that the overall process of going through larger and larger subRAFs is accurately reproduced by the model, and to confirm that in each repetition of the experiment (simulation), a different trajectory is indeed followed, as suggested by the Hasse diagram.

\subsection*{The advantage of RAF sets}

As a further demonstration of the value of the modeling approach, we consider the ``cooperation versus selfishness'' question \cite{Vaidya:12}. Using a variant of the formal model, we investigate the system of three cooperative molecules and three equivalent selfish ones, competing for the same food resources. The reaction graph of this system is shown in Fig. 5 (inset). Ribozymes E1 and S1 are assemblies of one pair of food molecules ($f_1$ and $f_2$), molecules E2 and S2 are assemblies of another pair of molecules ($f_3$ and $f_4$), and molecules E3 and S3 of yet another pair ($f_5$ and $f_6$). Molecule E1 catalyzes the assembly of E2, E2 that of E3, and E3 that of E1, in a cooperative way (forming an RAF set of size three). Molecules S1, S2, and S3 each catalyze their own assembly. So, the cooperative (green) part of the system (RAF set) competes with the selfish (red) part of the system for the same food molecules, as in the original experiment \cite{Vaidya:12}.

Fig. 5 shows a typical result of molecular flow simulations on this reaction network. Starting from a concentration of 100 (arbitrary units) of each of the six food molecules, the green line shows the concentration of E1+E2+E3 over time, and the red line that of S1+S2+S3 (the artificial dip around time point 2 will be explained below). Clearly, the RAF set (green) outcompetes the equivalent selfish system (red), and converts the majority of food molecules into the RAF molecules E1, E2, and E3. Note how similar this graph looks to the one from the original experiment (\cite{Vaidya:12} Fig. 2a, mixed system). In Fig. 5, the difference between cooperation and selfishness is even larger, although the difference is not always this large in each simulation run. Fig. 6 (top) shows the results of another simulation on the same reaction network where the difference is much smaller. In fact, occasionally the selfish (red) system actually outperforms the cooperative (green) one, an example of which is shown in Fig. 6 (bottom).

There is a simple explanation for the difference between cooperation and selfishness in this model. To get the cooperative system (RAF set) going, only one of the three (green) reactions has to happen ``spontaneously'', i.e., uncatalyzed; this is always possible, but at a lower rate (by a factor $c$) than a ribozyme-catalyzed reaction. This happens around time point 0.14 in the simulation shown in Fig. 5. However, in the selfish system, all three (red) reactions have to happen spontaneously at least once to get the full system going. This results in an almost three times longer waiting time on average. In the simulation shown in Fig. 5, the third spontaneous (uncatalyzed) red reaction happens around time point 0.35. By that time, however, the (green) RAF set has already built up enough ``momentum'' to outcompete the selfish (red) system, i.e., enough molecules of types E1, E2, and E3 are already around to increasingly catalyze each other's assembly. However, due to the stochastic nature of the system (having to wait for uncatalyzed reactions to happen), occasionally the selfish (red) system gets a head-start (with low probability) and outcompetes the cooperative (green) system.

Finally, there is also an advantage for the RAF set in terms of robustness. Suppose that at some point all molecules of type E3 and S3 are removed from the system, and a new supply of their food molecules ($f_5$ and $f_6$) is provided. This happens around time point 2 in Fig. 5, at the sudden dip in the concentrations. Since there are still E1 and E2 molecules present, the (green) RAF set quickly recovers without any delay. However, the (red) selfish system again has to wait until reaction S3 happens uncatalyzed at least once (which still has not happened by time point 3). Clearly, this suggests that RAF sets are more robust than selfish systems against perturbations, another prediction that can be tested with real chemical experiments.

\section*{Conclusions}

We have taken the chemical RNA replicator system described recently \cite{Vaidya:12} and formalized and investigated it within our mathematical RAF framework. As the experimental results showed, autocatalytic sets seem to form spontaneously in this chemical system \cite{Vaidya:12}. The existence of RAF sets in this system is verified by the formal model. In fact, many of the experimental results are accurately reproduced by the model, such as the emergence of larger and larger RAF sets over time, and the advantage of cooperative systems over selfish systems when they compete for the same resources. Of course there are many refinements and additional details that can be added to the current model, but even in its most basic form it already captures the main structural and dynamical properties of the real chemical system.

Moreover, the modeling approach provides several additional results and insights. First, it allows for a correction in the misinterpretation of some of the original experimental results. Second, it shows the ``richness'' of the RNA system in terms of the many subRAFs and ways they can be combined into larger subRAFs (as visualized by the Hasse diagram). Third, this ``richness'' suggests the testable prediction that each repetition of the experiment will, most likely, follow a different trajectory towards a realization of the maximal RAF set, which is confirmed computationally by the molecular flow simulations presented here. Lastly, it provides a simple explanation for the advantage of RAF sets over selfish systems, together with another testable prediction that RAF sets are more robust than selfish systems against perturbations.

In conclusion, we have shown that the formal RAF framework can be directly and meaningfully applied to real chemical systems, generating additional insights that are hard or even impossible to obtain from experiments alone, and thus provides a more detailed understanding of the nature of chemical systems in general and the emergence of autocatalytic sets in particular. This forms an important and much needed first step towards merging experimental and theoretical lines of work on autocatalytic sets in the context of the origin of life.

\section*{Competing interests}

The authors declare that they have no competing interests.

\section*{Authors' contributions}

W.H. and M.S. developed the RAF framework, W.H. constructed and analyzed the formal RNA replicator model, and W.H. and M.S. wrote the paper.

{\ifthenelse{\boolean{publ}}{\footnotesize}{\small}
\bibliographystyle{bmc_article}
\bibliography{RNA_RAF}}

\section*{}

\subsection*{Figure 1 - The corrected diagrams of \cite{Vaidya:12} Fig. 4.}
The existing subsets at two different time points in the original experiment, with the missing catalysis arrows included. To maintain consistency with previous work, catalysis arrows are shown with dashed lines. The nodes within the box in the left diagram (1h time point in the original experiments) form an RAF set of size seven. The diagram on the right (8h time point in the original experiments) forms an RAF set of size 11.

\vspace{1cm}
\includegraphics[scale=.6]{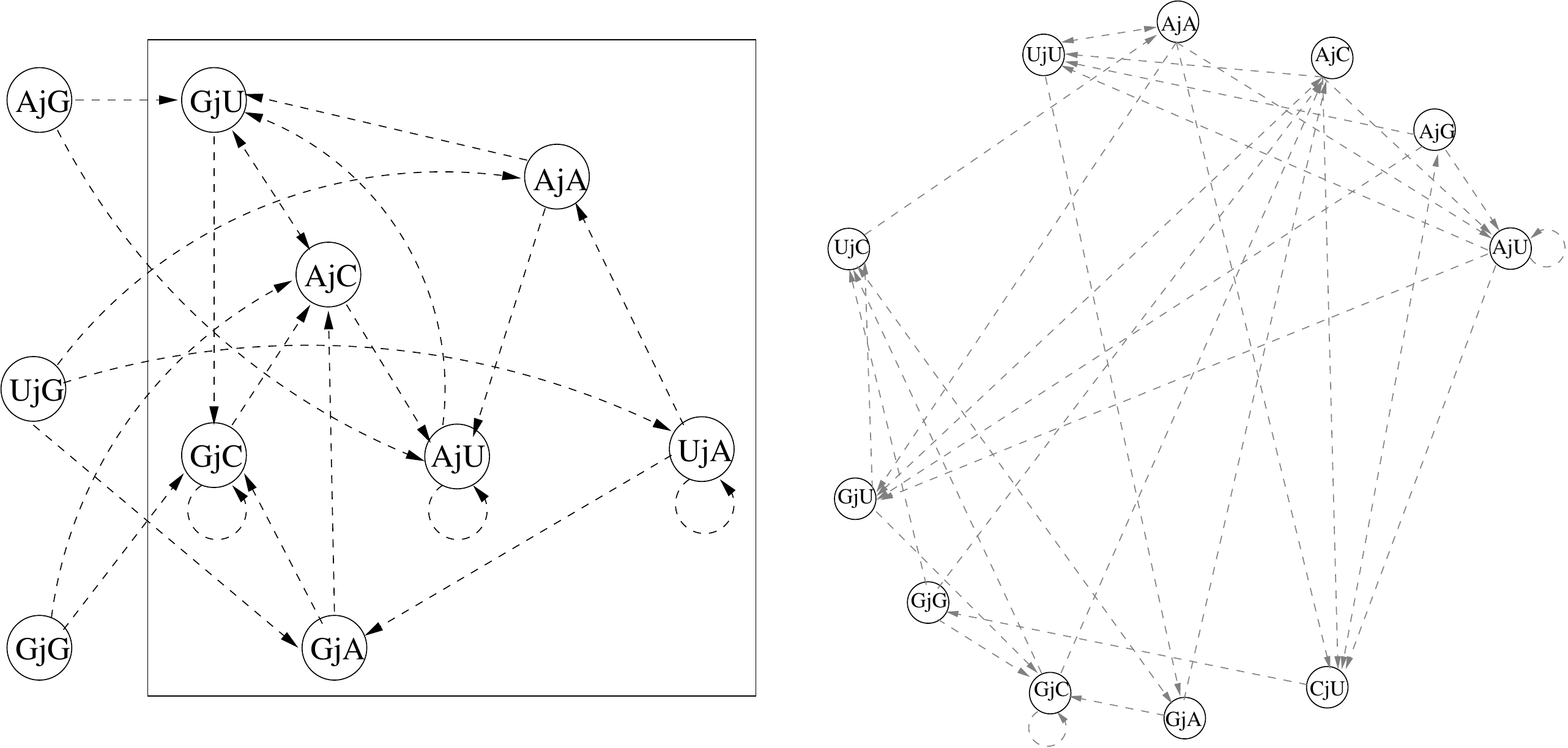}

\newpage

\subsection*{Figure 2 - The Hasse diagram of the subRAFs of the 7-reaction RAF set.}
The bottom row represents subRAFs of size one (the three autocatalytic reactions in Fig.1 (left)). Each next row up represents subRAFs of a larger size, up to the full 7-reaction RAF set at the top. Due to space constraints, only the four irreducible RAFs in this diagram (the four nodes that do not have an incoming edge from a lower level) and the full 7-reaction node at the top are labeled.

\vspace{1cm}
\includegraphics[scale=.6]{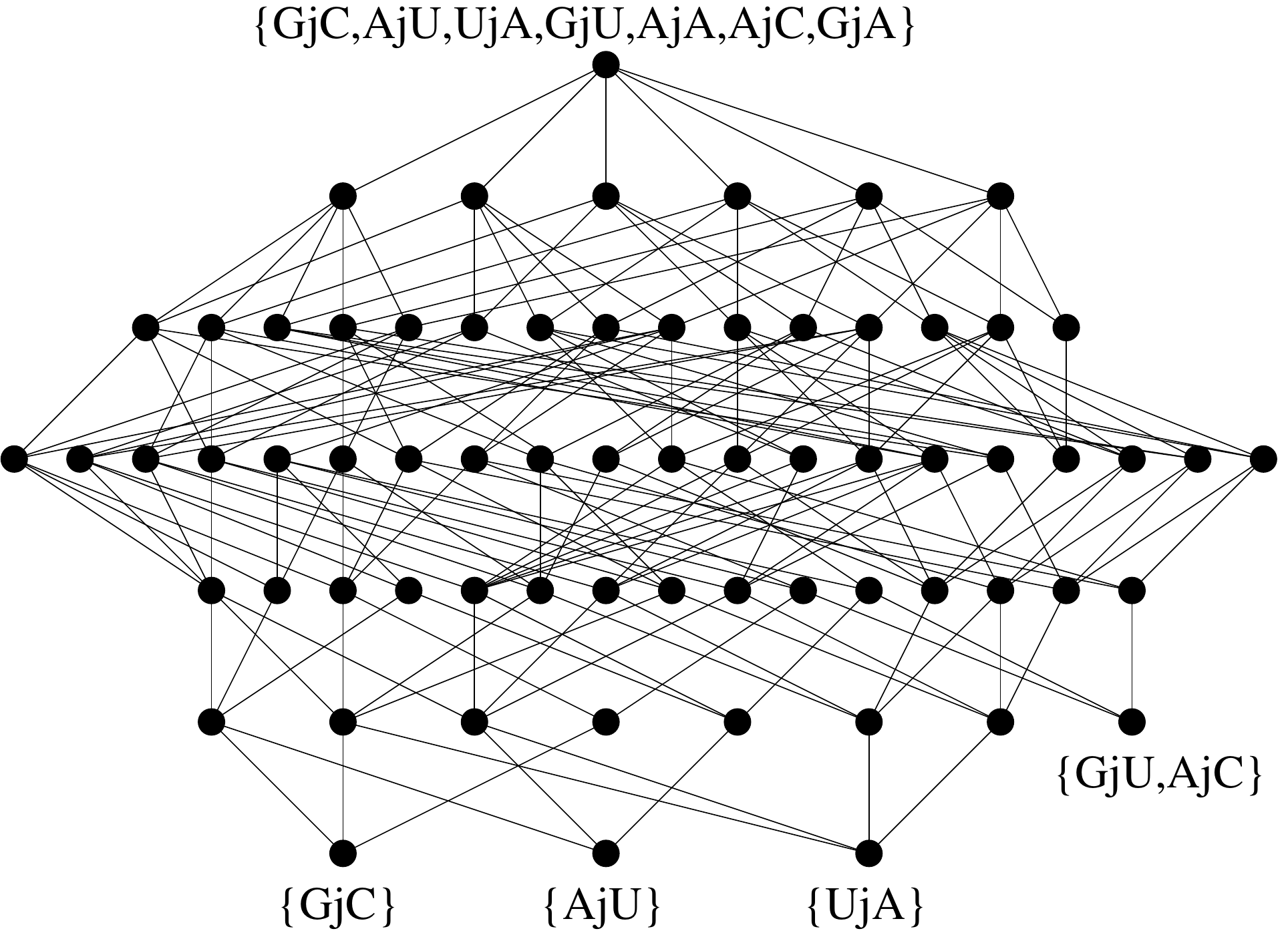}

\newpage

\subsection*{Figure 3 - The result of a molecular flow simulation on the RNA replicator model.}
The colored lines show the concentrations of the 16 ribozymes in the RNA replicator model over time, starting from only food molecules. Time and concentration are in arbitrary units.

\vspace{1cm}
\includegraphics[scale=.6]{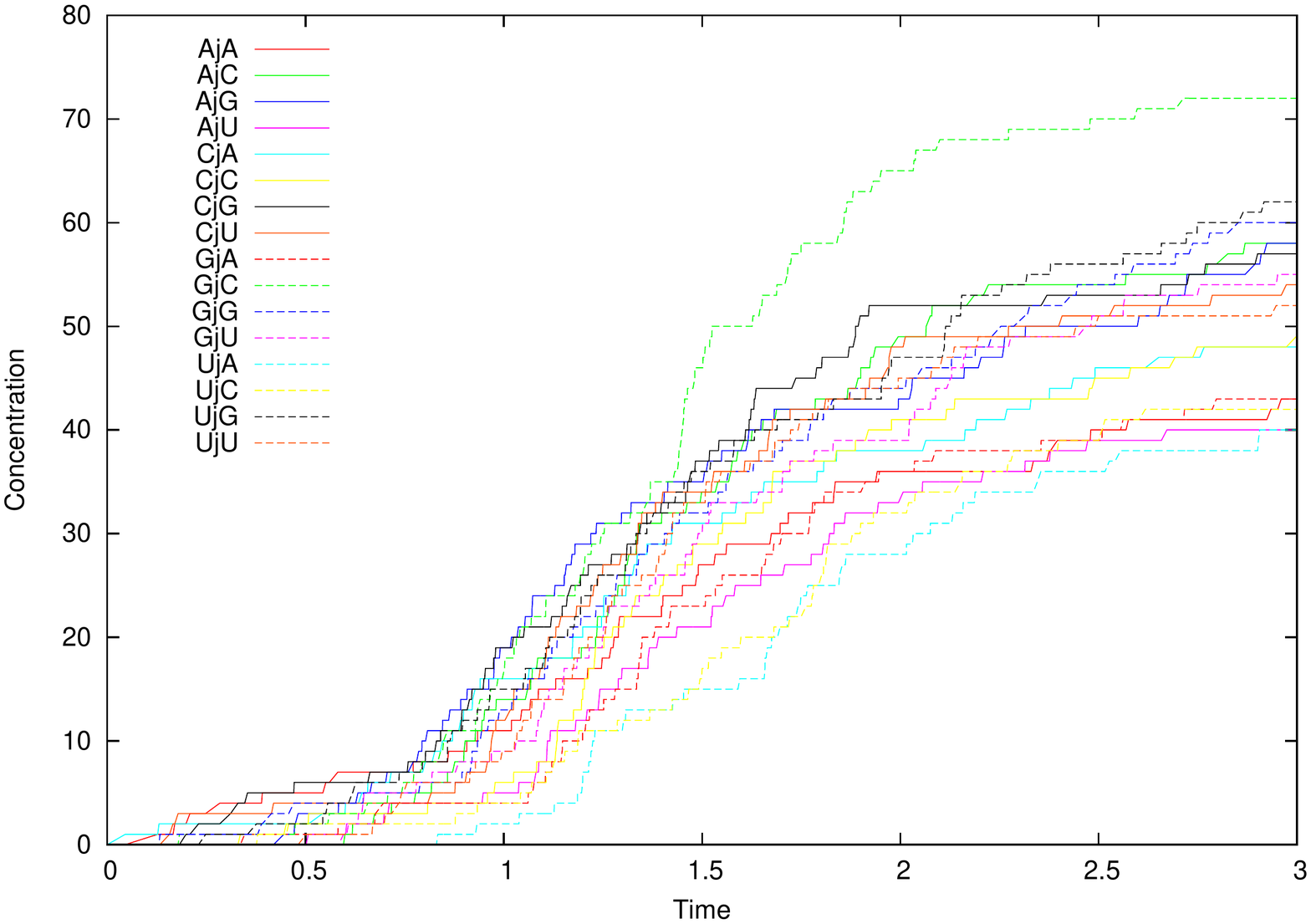}

\newpage

\subsection*{Figure 4 - The RAF subsets existing at various time points during the molecular flow simulation.}
The 16-reaction maximal RAF set is shown in grey. In each diagram, the subRAF existing in the system at some point in time is indicated in blue. The subsequent subRAFs shown are of size 1 (top left), 3 (top right), 7 (bottom left), and 11 (bottom right).

\vspace{1cm}
\includegraphics[scale=.7]{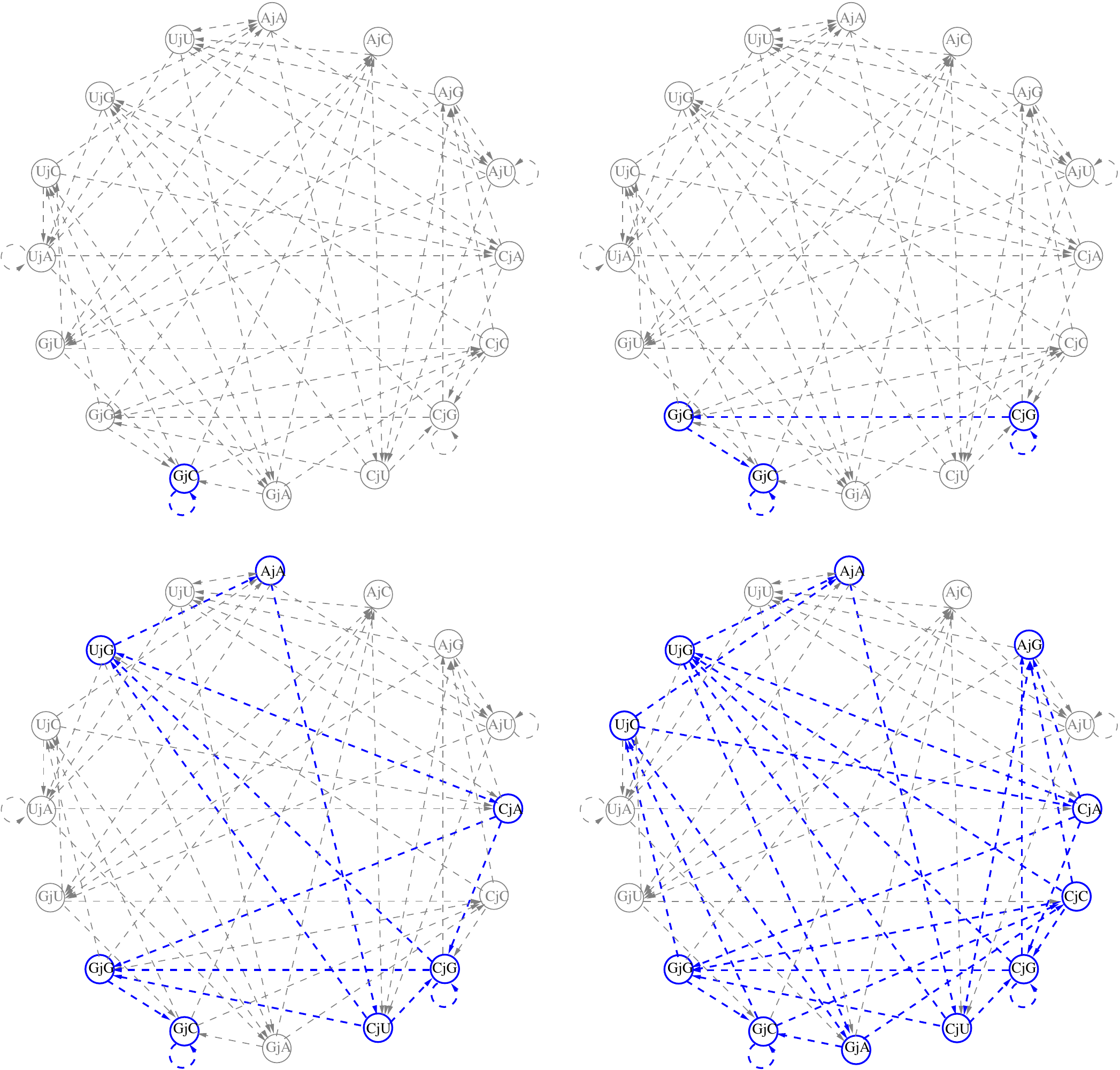}

\newpage

\subsection*{Figure 5 - Cooperation vs. selfishness.}
The cooperation (green nodes) and selfish (red nodes) reaction network is shown in the inset. The green nodes form an RAF set of size three. The plot shows the result of a molecular flow simulation on this reaction network. The green line shows the concentration of E1+E2+E2, while the red line shows the concentration of S1+S2+S3. At around time point 2, all molecules of type E3 and S3 are removed from the system, and a new supply of their food molecules is provided.

\vspace{1cm}
\includegraphics[scale=.6]{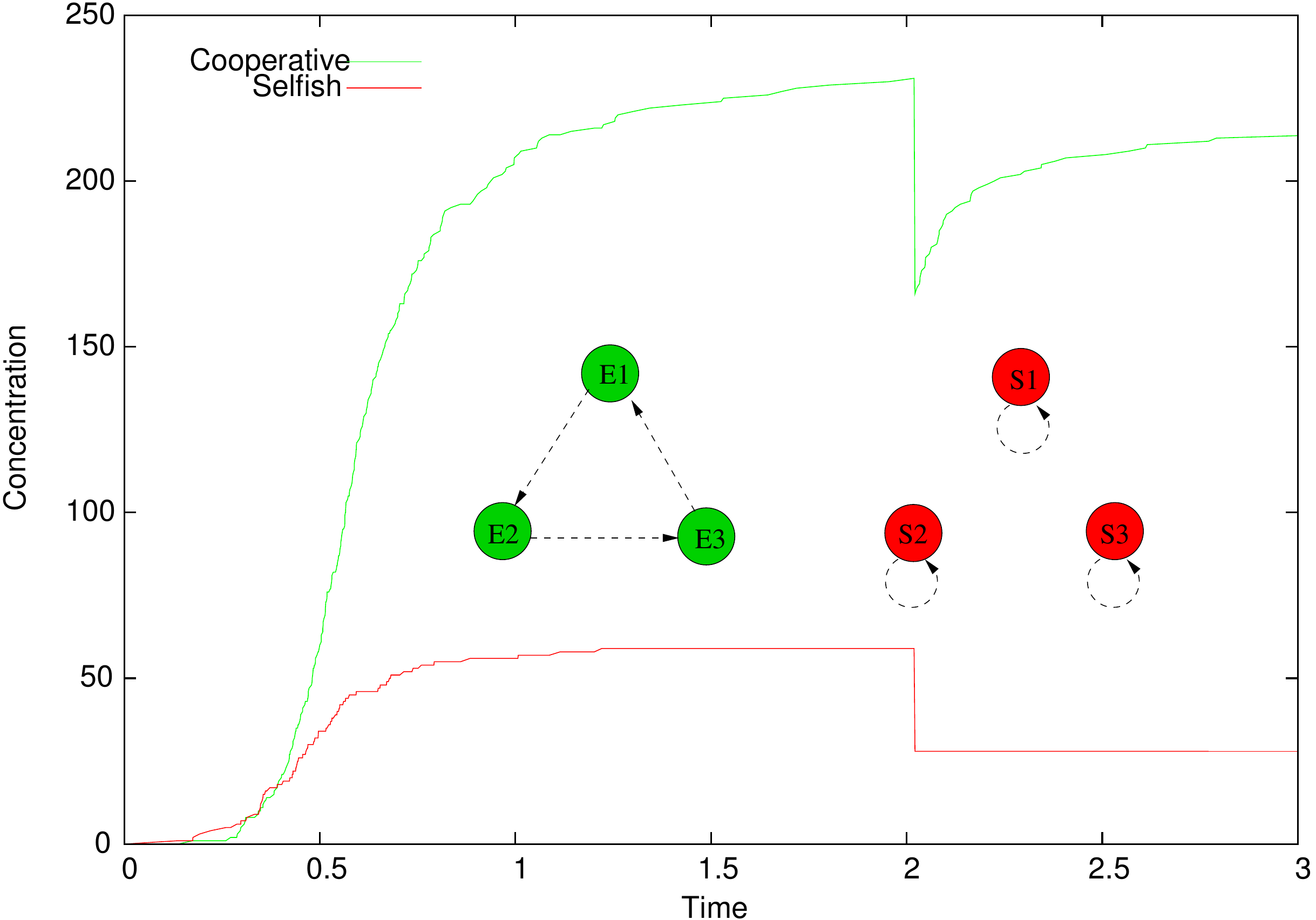}

\newpage

\subsection*{Figure 6 - Cooperation vs. selfishness, alternative results.}
Top: an example where the difference between the two systems is only minimal. Bottom: an example where the selfish system actually outcompetes the cooperative one.

\vspace{1cm}
\includegraphics[scale=.5]{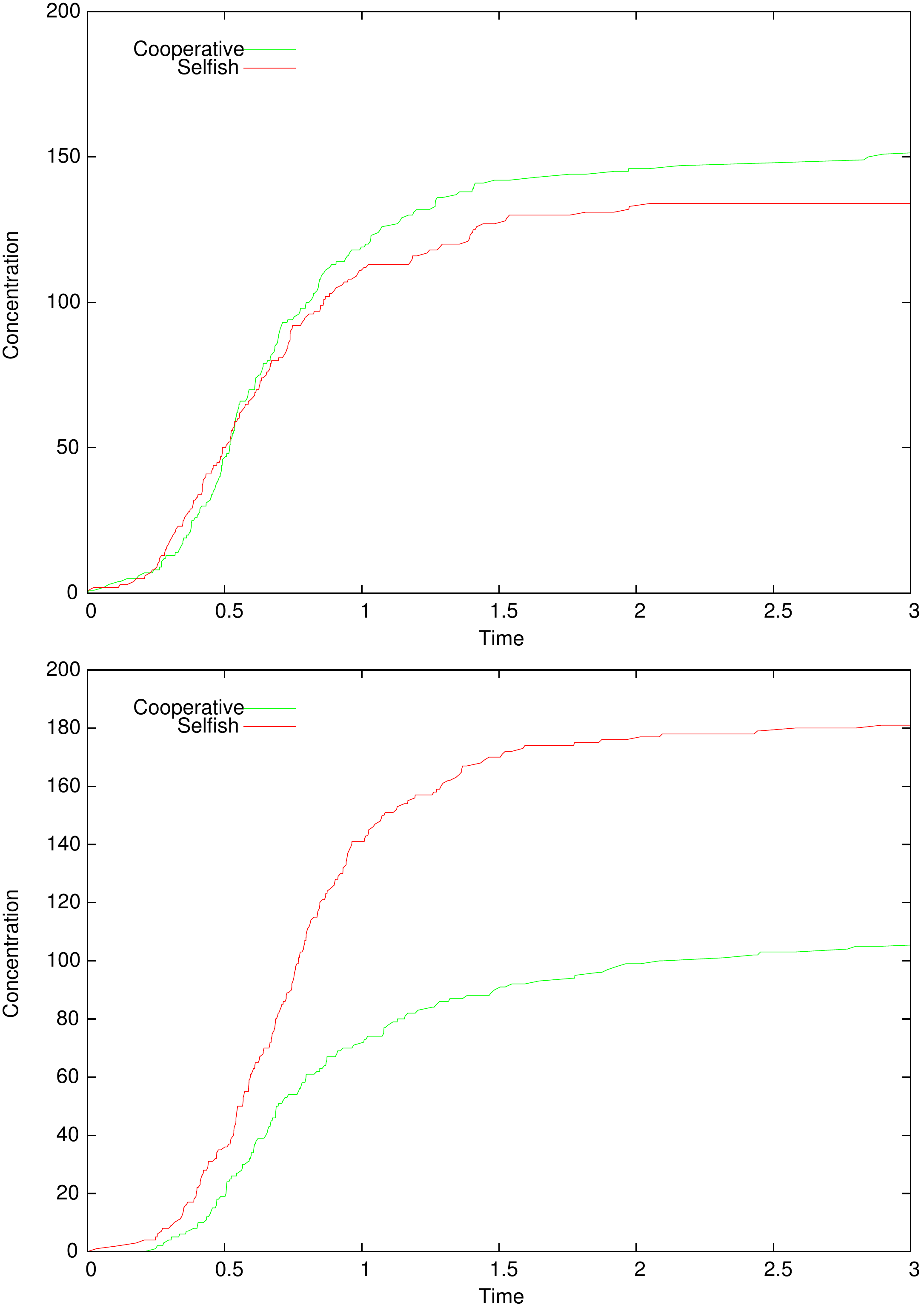}

\end{bmcformat}
\end{document}